\documentstyle[aps,prc,psfig]{revtex}            %PRC-Layout

\newcommand \be{\begin{eqnarray}}
\newcommand \ee{\end{eqnarray}}
\newcommand{\set}[2]{\newcommand{#1}{#2}}  %something else
\set{\pa}{\frac{\partial}{\partial}\, }
\begin{document}
\draft
\twocolumn[\hsize\textwidth\columnwidth\hsize %Erzeugt zweiseitiges Layout im
           \csname @twocolumnfalse\endcsname  %richtigen PRC-Format (prc s.o.)

%
%----------------Kopf des Paper's-------------------------------------------
%
\title{Damping Rates of Hot Giant Dipole Resonances}
%\title{Memory effects and collective excitations in hot nuclear matter}
\author{U. Fuhrmann, K. Morawetz$^*$, R. Walke$^*$}
\address{FB Physik, Universit\"at Rostock, D-18051 Rostock,
Germany\\
and $^*$ Laboratorio Nazionale Del Sud, Via S. Sofia, 44-95123
Catania, Italy}
\date{\today}
\maketitle
\begin{abstract}
The damping rate of hot giant dipole resonances (GDR) is investigated.
Besides Landau damping we consider collisions and density
fluctuations as contributions to the damping of GDR.
Within the non-equilibrium Green's function method we derive a non-Markovian
kinetic equation. The linearization of the latter one
leads to complex dispersion relations. The complex solution provides 
the centroid energy and the damping width of giant resonances. 
The experimental damping widths are the full width half maximum (FWHM) and 
can be reproduced by the full width of the structure function.           
Within simple finite size scaling we
give a relation between the minimal interaction
strength which is required for a collective oscillation and
the clustersize. We investigate the damping of giant dipole
resonances within a Skyrme type of interaction.
Different collision integrals are compared 
with each other in order to incorporate correlations. The inclusion of
a conserving relaxation time approximation allows to find the
$T^2$-dependence of the damping rate with a temperature known 
from the Fermi-liquid theory. 
However, memory effects turn out to be essential for
a proper treatment of the damping of collective modes.
We derive a Landau like formula for the one--particle relaxation 
time similar to the damping of
zero sound. 
\end{abstract}
\vskip2pc]

%
%***************************EINLEITUNG*****************************************
%
\section{Introduction}

Giant resonances are high frequency, collective excitation modes of a nucleus.
They can be identified as the collective motion in the nuclear volume
and are found as a property of all nuclei. In particular in the last years
the experimental and theoretical interest is focused to understand the
width of such 
giant 
resonances \cite{SBT91,ABA95,KPS95,BAB95,SUO96,MOT96,KPS96,TKL97,KTO96,HNP96}.
Although new experimental data are available at high excitation energy
\cite{RAP96,RAM96} the
theoretical description of the temperature dependence of damping rate
is still a matter of discussion.

The theoretical treatment of giant resonances can be roughly characterized 
by two approaches.
The first one considers the finite nucleus and solves 
the random phase approximation
(RPA) equation by
diagonalization of the Hamiltonian \cite{SWA91,SRS97}. In this approach the
damping width can be extracted as an envelope of discrete excitation
lines, sometimes called Landau fragmentation. However the introduction 
of the temperature remains difficult.
The second class of approaches relies on the high density of states and
consequently uses a continuous model, mostly the
Fermi liquid theory \cite{KPS95,KPS96,TKL97,KTO96,ATS94,KAM69,KLT97}.
Within this treatment, dispersion relations
are derived whose solutions provide the energy and the width of
collective excitations \cite{TKL97,KTO96,KLT97,HNP96}.

The microscopic theory is mainly based on
Vlasov kinetic equations \cite{BRI88,BDT86,DMA86}.
The influence of correlations by particle-particle collisions 
is investigated using
numerical solutions of Boltzmann--Uehling--Uhlenbeck (BUU)--type
equations \cite{BAR96,CTG93}.
To get more physical insight into these simulation results, collective
models based on the scaling theory are developed \cite{RSC82}.
It turns out that the non-Markovian kinetic equation is necessary to get realistic
values for giant monopole resonances \cite{ABA95}.
These collective models calculate
the damping rate by an average procedure of the collision integral \cite{BBB83}.

We follow another line. 
We will start from different kinetic equations and
derive dispersion relations for the collective modes by linearization of
the corresponding kinetic equation. Instead of collective averaging, 
we solve these
dispersion relations and obtain directly the influence of correlations on the
damping rate.
We present in the paper different
contributions to the damping of giant dipole resonances in a systematic way.

Let us shortly outline our theoretical approach to collective resonances.
The collective density oscillations are determined via the time dependence
of the one-particle distribution function $f(R,p,T)$ where the density reads
$n(R,T)=\int \frac{d^3p}{(2\pi)^3}\,f(R,p,T)$. The one-particle distribution
obeys the kinetic equation
\be
\dot{f}(R,p,T)&-&\{{p^2 \over 2m} +U(R,T)+U_{\rm
              ext}(R,T),f(R,p,T)\}\nonumber\\
&=&I_c[f(R,p,T)]\, ,\label{kinpoisson}
\ee
where $U$ is the mean-field potential and $I_c$ corresponds to the
collisional term. The Poisson brackets are abbreviated as
$\{a,b\}=\partial_Ra\partial_pb-\partial_pa\partial_Rb$. 
To get the linear response of the system to an external
field $U_{ext}$ we linearize equation (\ref{kinpoisson}) around a
quasi--equilibrium $f(R,p,T)=f^0(p)+\delta f(R,p,T)$
and get after Fourier transformation $T\rightarrow \omega$
and $R\rightarrow q$
\be
&&i \omega\delta f(q,p,\omega)-i {p q\over m} \delta f(q,p,\omega)
\nonumber\\
&+&i (U'[n] \delta n(q,\omega) +U_{\rm ex}) q \partial_p
f_0(p)=I_c[\delta f(q,p,\omega)].
\label{kinpoisson_linear}
\ee
Integrating over $p$ the solution provides the
polarization function $\Pi(q,\omega)$
\be
\Pi(q,\omega)=\frac{\delta n(q,\omega)}{U_{ext}(q,\omega)}
 \label{def_response}
\ee
with $\delta n(q,\omega)=\int \frac{d^3p}{(2\pi)^3}\,\delta
f(q,p,\omega)$.
In the RPA approximation which corresponds to the neglect of
collisions we obtain
\be
\Pi_{\rm RPA}(q,\omega)=\frac{\Pi^0(q,\omega)}{1-V_0\Pi^0(q,\omega)}\,
\label{def_response1}
\ee
with the standard form of polarization function $\Pi^0$
(\ref{Lindhard_polarisation}) and $V_0=\partial_n U[n]$.
The polarization function contains information
about the collective excitation properties of nuclear matter.
According to the denominator of (\ref{def_response1})
the relation to the dielectric function (DF) is given by  
\be
\epsilon(q,\omega)=1-V_0(n_0)\Pi^0(q,\omega) \, .\label{Def_DF}
\ee
For different collision integrals we will get different
polarization functions. The complex zeros $\omega_0=\Omega-i \gamma$
of (\ref{Def_DF}) determine the energy $\Omega$ and 
width $\gamma$ of the collective excitation.
>From the DF one has the spectral function (structure function) via
\be
{\rm Im}\frac{1}{\epsilon(q,\omega)}=-\frac{{\rm Im}\epsilon(q,\omega)}
                     {\big[{\rm Re}\epsilon(q,\omega)\big]^2+
                      \big[{\rm Im}\epsilon(q,\omega)\big]^2}\, ,
                       \label{inverseps}
\ee
which is important because it's structure reflects the collective excitations.
Sum rules, e.g. \cite{PNO68}
\be
\frac{1}{\pi V_0}\int\limits_0^{\infty}d\omega~\omega
{\rm Im}\frac{1}{\epsilon(q,\omega)}=\frac{n_0q^2}{2m}\,
                                           \label{f_summenregel}
\ee
are an exact property of the spectral function.

The paper is organized as follows. In Sec.\ref{kinetic} we start with
a generalized quantum kinetic equation which we rederived
in appendix \ref{GKEq} using the Martin-Schwinger hierarchy for the 
real-time Green's function.
To include memory effects into the kinetic equation we use the
generalized Kadanoff-Baym Ansatz and obtain a non-Markovian kinetic
equation.
In Sec.\ref{dielectric} we linearize these kinetic equations
to get the DF for infinite hot nuclear matter in four different
approximations: (Sec.\ref{vlasovU}) collisionless Vlasov equation,
(Sec.\ref{BUU}) conserving relaxation time
approximation (Mermin approximation), (Sec.\ref{nonMarkov})
dynamical relaxation time approximation (reflecting
memory effects) and (Sec. \ref{fluk}) the effect of density fluctuations on the
potential.
For these approximations we compare 
the damping rates and centroid energies of GDR which are
the complex solution of the dispersion relation.  
In Sec.\ref{exper} we discuss these results together with the
FWHM of the structure function and with experimental data.

%
%**********************KINETISCHE GLEICHUNGEN**********************************
%

\section{Kinetic equation approach}
\label{kinetic}

Let us start from the kinetic equation in general 
form (\ref{akbE1}) from appendix \ref{GKEq} 
\be
\lefteqn{\left[\frac{\partial}{\partial T}+\frac{p}{m}\cdot\nabla_R\right]
f_W(p,R,T)}\hspace{1cm}                                             \nonumber\\
&+&\int dr\frac{dp'}{(2\pi)^3}\,e^{i(p'-p)\cdot r}f_W(p^\prime,R,T) \nonumber\\
&&\times\left[\Sigma_H(R+\frac{r}{2},T)-\Sigma_H(R-\frac{r}{2},T)\right]=I_c,
                                                          \label{hartreeG}   \\
\lefteqn{I_c=\int\limits_{-\infty}^{0}d\tau\left[
                \left\{G^>(p,R,T-\frac{\tau}{2},\tau),
 \Sigma^<(p,R,T-\frac{\tau}{2},-\tau)\right\} \right.} \hspace{1cm} \nonumber\\
     &&\left.-\left\{G^<(p,R,T-\frac{\tau}{2},\tau),                
              \Sigma^>(p,R,T-\frac{\tau}{2},-\tau)\right\}  \right],\nonumber\\
                                                                 \label{allgI}
\ee
where the Wigner
distribution function $f_W$ is connected to the correlation function
$f_W(p,R,T)=G^<(p,R,T,\tau=0)$. Further $\{,\}$ is the anti-commutator 
of integrals over Wigner coordinates.\\
Neglecting the collision integral $I_c$ on the right hand 
side of (\ref{hartreeG}) one obtains
the collisionless quantum Vlasov equation \cite{KBA62}. This leads to
the Lindhard polarization function (\ref{Lindhard_polarisation}).\\
Next, we will consider binary collisions and will use for the self-energy
in (\ref{allgI}) the 1. Born approximation
\be
\Sigma_{C}^{\stackrel{>}{<}}(p,t-t^\prime)&=&
\int\frac{ d^3p^\prime d^3{\bar p}\,d^3{\bar p}^\prime} {(2\pi)^9} \\
&\times&W(pp^\prime,{\bar p}{\bar p}^\prime)(2\pi)^3
\delta(p+p^\prime-{\bar p}-{\bar p}^\prime)        \nonumber\\
&\times&G^{\stackrel{<}{>}}(p^\prime,t^\prime-t)
         G^{\stackrel{>}{<}}({\bar p},t-t^\prime)
         G^{\stackrel{>}{<}}({\bar p}^\prime,t-t^\prime) \, ,\nonumber
\ee
where $W$ is the collision probability. We close equation (\ref{hartreeG})
applying the generalized Kadanoff-Baym ansatz \cite{LSV86}
\be
 G^<(p,R,T,\tau)=e^{-\frac{i}{\hbar}\int\limits_{T-{\tau \over 2}}^{T+{\tau \over
2}} dt\epsilon(p,R,t)}
                f_W(p,R,T-\frac{|\tau|}{2})\,  \label{akbansatz}
\ee
which gives a
connection between the correlation functions $G^{\stackrel{>}{<}}$ and
the Wigner distribution.
The quasi-particle energy $\epsilon(p,R,T)$ in the quasi-particle
picture is given by the solution of the dispersion relation
\be
\epsilon=\frac{p^2}{2m}+U(R,T)+{\rm R}e\Sigma(p,\epsilon,R,T)\label{e_dispersion}.
\ee

The resulting non-Markovian collision integral reads now
\be
&&I_{m}(p_1,T)=\int\limits_{0}^{\infty}d\tau\int \frac{d^3p_2 d^3p_3} {(2\pi)^6}
\,W(p_1p_2,p_3p_4)\nonumber \\
&\times&\left [ U^+(T-\tau,T)+U^-(T-\tau,T) \right ]\nonumber \\
&\times&\left[f_3(T-\tau)
f_4(T-\tau)(1-f_1(T-\tau))(1-f_2(T-\tau))\right .\nonumber \\
&-&\left .f_1(T-\tau)f_2(T-\tau)(1-f_3(T-\tau))(1-f_4(T-\tau))\right]
\nonumber\\
\label{memorystoss1}
\ee
where $f_i(T)=f(p_i,R,T)$ ($i=1,2,3,4$), $p_4=p_1+p_2-p_3$ and the
full time dependent propagator $U^\pm$ is 
\be
U^\pm(T-\tau,T)=
\exp[\pm i \int\limits_T^{T-\tau}dt^\prime
\bigtriangleup\epsilon(t^\prime)].
\ee
$\bigtriangleup\epsilon$ are the time dependent
quasi-particle energies
$\bigtriangleup\epsilon(t^\prime)=
\epsilon_1(t^\prime)+\epsilon_2(t^\prime)-\epsilon_3(t^\prime)-
\epsilon_4(t^\prime)$.
If memory effects are neglected, equation(\ref{memorystoss1}) becomes the usual
Boltzmann or BUU collision integral $I_B$
\be
I_B&=&\int \frac{d^3p_2 d^3p_3} {(2\pi)^6}
\,W(p_1p_2,p_3p_4)\nonumber\\
&\times&\delta(\epsilon(p_1)+\epsilon(p_2)-\epsilon(p_3)-\epsilon(p_4) \nonumber\\
&\times&\left[f_3 f_4(1-f_1)(1-f_2)-f_1f_2(1-f_3)(1-f_4)\right]\, .
\label{boltzmannstoss1}
\ee

The Boltzmann collision integral is modified in (\ref{memorystoss1})
by a broadening of the delta distribution function of the energy
conservation and an additional retardation in the center-of-mass times
of the distribution functions $f_i=f(p_i,R,T-\tau)$. The first effect is
connected with phase decay or spectral properties
and responsible for global energy conservation \cite{M94,MSL97}. The second effect
gives rise to genuine memory effects.
The formation of
correlations at such short time scales are discussed in \cite{MSL97}.

%
%***********************ERGEBNISSE*******************************************
%

\section{Collective excitation and dielectric function}
\label{dielectric}

We now linearize the different derived kinetic
equations and analyze their consequences to the damping of GDR.
The standard RPA is first repeated in order to explain our analysis.

\subsection{Vlasov equation-collisionless Landau damping}
\label{vlasovU}
The linearization of the quantum Vlasov equation (\ref{hartreeG}) yields the
RPA of the dielectric function which has the form of equation (\ref{Def_DF}),
but $\Pi_0$ is the (complex) Lindhard polarization function \cite{LIN54}
\be
\Pi_0(\Omega,\gamma,q)=2\,\int_c\frac{d^3k}{(2\pi)^3}
    \frac{f(k)-f(k+q)}{E_{k}-E_{k+q}+(\Omega-i\gamma)+i\eta}
    \label{Lindhard_polarisation}
\ee
where $E_k=k^2/2m$ and $\eta$ is an infinitesimal small number.
Spin degeneracy has been accounted for.
$\Omega$  and $\gamma$ denote, respectively, the real part and the
negative of the imaginary part of the frequency ($\omega=\Omega-i\gamma$).
By consideration of a simplified Skyrme force \cite{BRV94}
\be
\upsilon^{'}&=&t_0(1+x_0\hat{P}_\sigma)\delta(r_1-r_2)+
             t_3\delta(r_1-r_2)\delta(r_2-r_3)\, ,\nonumber\\
&&
\label{Skyrme}
\ee
one obtains a mean-field potential for the neutrons
$U_n$ \cite{VBR72,BRV94}
\be
U_n(R,T)&=&t_0\left\{\left(1+\frac{x_0}{2}\right)(n_n(R,T)+n_p(R,T))\right
.\nonumber\\
&-&\left . \left(x_0+\frac{1}{2}\right)n_n(R,T)\right\}
\nonumber\\
&+&
\frac{t_3}{4}\left\{(n_n(R,T)+n_p(R,T))^2-n_n^2(R,T)\right\}\label{PotNeu}
\ee
and $U_p$ is given by an interchange of $n_p$ and $n_n$.
>From (\ref{kinpoisson_linear})
and (\ref{def_response})
we can read off the effective particle-hole potential $V_0$ for isovector modes
\be
U_n-U_p&=& 2 V_0\left[\delta n_n-\delta n_p\right]\nonumber \\
V_0(n_0)&=&-\frac{t_0}{2}\left(x_0+\frac{1}{2}\right)-\frac{t_3}{8}(n_n+n_p)\,.
\label{Potential_V0}
\ee
The parameter $t_0$, $t_3$ and $x_0$ were fitted to
reproduce the binding energy ($E/A=-16$ MeV) at
the saturation density ($n_0=0.17$ fm$^{-3}$) of nuclear matter.
For the GDR the wave vector $q$ is estimated according to the formula
\cite{VBR72,SJE50}
\be
q=\frac{\pi}{2\,R}\, .
\label{scalingL}
\ee
In this model neutrons and protons oscillate out of phase
inside a sphere of the radius $R$ (=nuclear radius). Focusing our
interest to the nucleus $^{208}$Pb ($^{120}$Sn)  we use
$q\approx 0.23$ fm$^{-1}$ ($q\approx 0.277$ fm$^{-1}$)
with $R=6.7$ fm ($R=5.6$ fm).

The dispersion of collective excitation is now computed from the zeros
of the complex dielectric function (DF), Eq. (\ref{Def_DF}),
\be
\mbox{Re}\,\epsilon(\Omega-i \gamma,q)+\mbox{Im}\,\epsilon(\Omega
-i \gamma,q)=0 \label{complexdisp}
\ee
where $\gamma$ gives the Landau damping of the collective excitation.
An approximate solution of this RPA dispersion relation is possible
if the damping  Im$\epsilon$   is small \cite{KKE86}. Then one can
linearize the collective excitation spectrum
\be
&&{\rm Re} \epsilon(q,\Omega)+(\omega-\Omega+i\gamma) \partial_{\omega} 
{\rm Re} \epsilon(q,\Omega) +i {\rm Im} \epsilon(q,\Omega)=0\nonumber\\
&&
\ee
which leads to
\be
\gamma={{\rm Im} \epsilon(q,\Omega) \over \partial_{\Omega} {\rm Re}
\epsilon(q,\Omega)}
\ee
and $\Omega$ the solution of ${\rm Re} \epsilon(q,\Omega)=0$.
This is, however only justified for small values of the damping $\gamma$.
The correct procedure is to
carry out the analytical continuation of the DF into the lower energy plane.
Performing the integration one can express the DF
(\ref{Def_DF}) with  (\ref{Lindhard_polarisation})
by the dimensionless variables
$x=q/\sqrt{2mT}$, $z^{\ast}=\Omega/2T$ and $\xi=-\gamma/2T$ in the form
\be
\epsilon(x,z^{\ast},\xi)&=&1-\frac{V_0 c}{2x}
\int\limits_{C}dz\frac{F(z)}{z-z^{\ast}+i\xi} \quad \mbox{where}
\label{DFzylinder}\\
F(z)&=&\ln\frac{1+\varsigma \exp-\left(z-\frac{x}{2}\right)^2}
           {1+\varsigma \exp-\left(z+\frac{x}{2}\right)^2}\, ,\label{F_func}
\ee
$c=\sqrt{2 T m^3}/\pi^2$ and $\varsigma=e^{\mu/T}$ denotes the fugacity.
Following Landau's contour integration \cite{LPI79}
the result of the analytic continuation of the DF (\ref{DFzylinder})
with the pole $z_0=z^{\ast}-i\xi$ is
\be
&&\epsilon^c(x,z^{\ast},\xi)=\nonumber\\
&&1-\frac{V_0 c}{2x}
 \left\{
  \begin{array}{ll}
    {\displaystyle\wp\int\limits_{-\infty}^{+\infty}dz
           \frac{F(z)}{z-z^{\ast}}+i\pi F(z^{\ast})}   & \xi=0 \\
    {\displaystyle\int\limits_{-\infty}^{+\infty}dz
   \frac{F(z)}{z-z^{\ast}+i\xi}+i2\pi F(z^{\ast}-i\xi)}& \xi>0 \\
    {\displaystyle\int\limits_{-\infty}^{+\infty}dz
                   \frac{F(z)}{z-z^{\ast}+i\xi}}       & \xi<0.
  \end{array} \right.  \nonumber\\
&& \label{landauDF}
\ee
The explicit expressions of the real and imaginary parts of the
analytical continuation of the DF are given in appendix \ref{herleitungenA}.
For complex values of $z$ there are also poles of
the function $F(z)$ in (\ref{F_func})
which require a separate investigation. They are located at
\be
z_n=\pm\frac{x}{2}\pm\sqrt{\frac{\vartheta_n}{T}}\, ,
  \label{matsubara}
\ee
where $\vartheta_n$ are the discrete fermionic Matsubara frequencies
\be
\vartheta_n=\mu-i(2n+1)\pi T\,;\qquad n=0,\pm 1,\pm 2,... \,\,.
\ee
If these poles do not agree with the poles $z_0$ of the
denominator of equation (\ref{DFzylinder}) there is
no contribution to the integration
due to the fact that {\tt Res}$[F(z)/(z-z_0);z_n]=0$. Because of
the infinite residue the remaining
case $z_0= z_n$ is found to be singular and will be discussed elsewhere.

Fig. \ref{crossT0_paper} shows the zeros of the real and the imaginary part
of the DF equation (\ref{landauDF}) in the complex plane 
at zero temperature (left hand side). The right hand side
represents  their position in the pair excitation spectrum which
is bounded by the lines $\omega_\pm=q/2m(q\pm 2 p_F)$.
For increasing wave vectors $q$ we have found three special cases:  
%--------------------------Fig. 1 --------------------------------------------

\begin{figure}
\centerline{\psfig{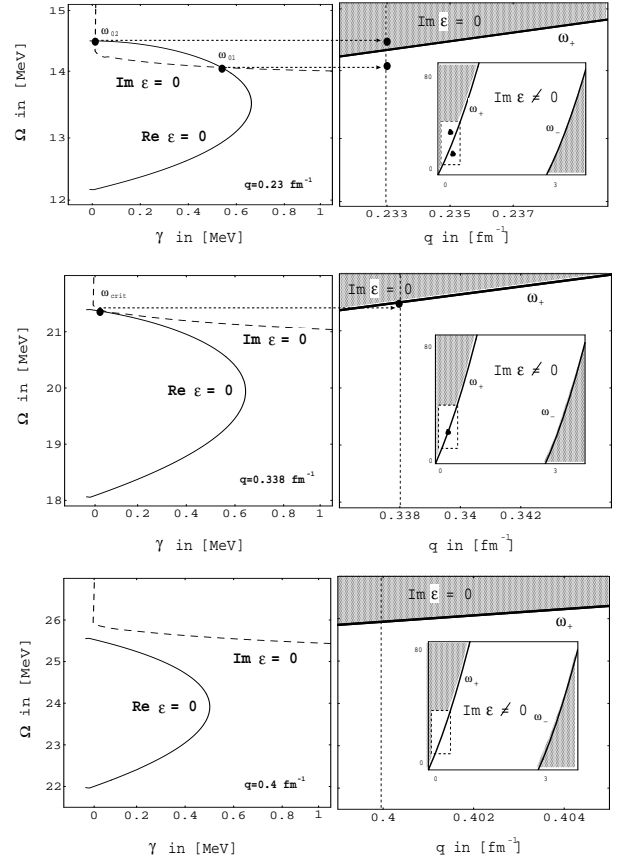}}
\vspace{.5cm}
\caption{Zeros of the analytic continuation of the complex DF (Lindhard)
          for different wave vectors at $T=0$. LHS:
          collective excitations (GDR) in nuclear matter
          correspond to the crossing of the lines
          Re $\epsilon=0$ and Im $\epsilon=0$ (marked by thick dots). RHS:
          pair continuum  Im $\epsilon \neq 0$ and
          undamped (marked) region Im $\epsilon=0, \gamma=0$ with
          $\omega_\pm=q/2m(q\pm 2 p_F)$. }
                      \label{crossT0_paper}
\end{figure}
{\bf Top} (q=0.23 fm$^{-1}$): There are two collective excitations
which correspond
to the crossing points of the zeros of Re($\epsilon^c$) and Im($\epsilon^c$).
Their resolution in the pair continuum (RHS) yields an undamped collective
excitation $\omega_{02}$ which lies outside the pair continuum (marked region,
Im$(\epsilon)=0$, $\gamma=0$). The collective excitation $\omega_{01}$ lies
inside the pair continuum and is therefore damped (Im$(\epsilon)\neq 0$,
$\gamma\neq 0$). From the DF one can now calculate the spectral function
(\ref{inverseps}). In Fig. \ref{inversEpsT0_paper} one recognizes
a $\delta$-shaped peak at the frequency of the
undamped excitation $\omega_{02}$ and a broader peak corresponding to
the damped collective excitation $\omega_{01}$.\\
{\bf Middle} (q=0.338 fm$^{-1}$): There arises only one critical
collective excitation $\omega_{crit}$ for a critical wave vector $q_{crit}$.
The collective excitation $\omega_{crit}$ lies inside the
pair continuum ($\omega_{crit}\approx \omega_+$) and  is damped. \protect\\
{\bf Bottom} (q=0.4 fm$^{-1}$): Since there is no crossing of the zeros
of the real part and the imaginary part of the DF, no collective
excitation can occur.

%%-----------------------Fig. 2. --------------------------------------------
\begin{figure}
\centerline{\psfig{file=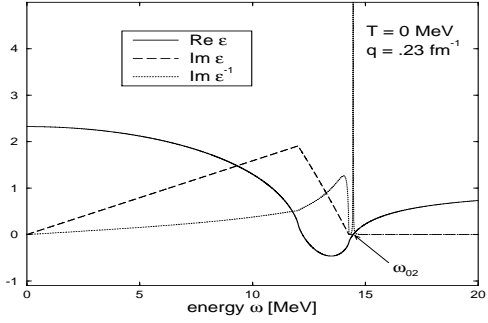,height=5.cm,width=8cm,angle=-90}}
\vspace{.5cm}
 \caption{Real part (solid line), imaginary part (dashed line) and
          spectral function (dotted line) of the Lindhard DF
          for the wave vector $q=0.23$ fm$^{-1}$ at $T=0$ corresponding to
          Fig. \ref{crossT0_paper} ({\bf top}). }
                      \label{inversEpsT0_paper}
\end{figure}

\begin{figure}
\centerline{\psfig{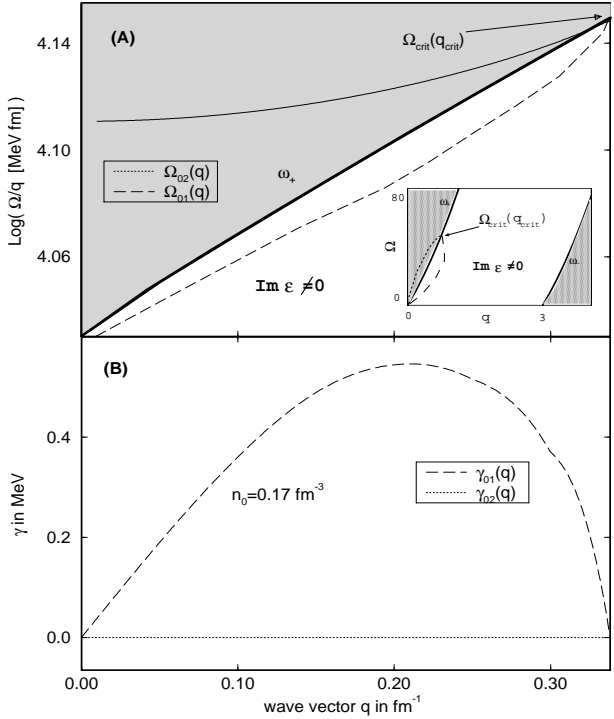}}
\vspace{.5cm}
 \caption{{\bf (A)} Dispersion of the collective excitations at $T=0$
    of Fig. \ref{crossT0_paper} and their position in the
    pair continuum. The mode $\Omega_{02}(q)$ (dotted line) is undamped and
    $\Omega_{01}(q)$ (dashed line) is damped. The insert shows an
    enlarged view of the pair continuum where the marked region corresponds
    to the undamped region (Im=0).
    {\bf (B)} Corresponding damping rates. }
                      \label{dispersionT0_paper}
\end{figure}

In Fig. \ref{dispersionT0_paper} we summarize the above results 
considering the entire
dispersion of the collective excitation at $T=0$. We see that
nuclear matter at zero temperature has a region where two collective modes
are excited (Fig. \ref{dispersionT0_paper} ({\bf A})).
The mode $\Omega_{02}$ goes outside the pair continuum and is undamped,
whereas the mode $\Omega_{01}$ propagates inside as a damped one.
%--------------------------Fig. 3 --------------------------------------------

\begin{figure}
\centerline{\psfig{file=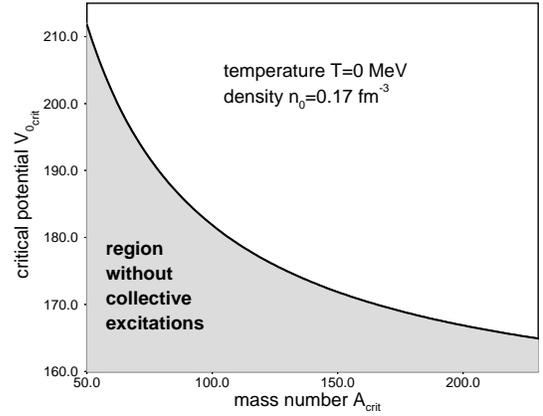,height=6.5cm,width=8cm,angle=-90}}
\vspace{.5cm}
 \caption{According to Fig. \ref{crossT0_paper} (middle) calculated
          critical potentials $V_{0_{crit}}$ for critical
          wave vectors $q_{crit}$ which are connected with
          the mass number $A$ via (\protect\ref{scalingL}).
          The marked area corresponds to a region where no collective 
          excitations exist.}
                      \label{AkritvsV0krit_paper}
\end{figure}

Beyond the coincidence of both modes into the critical point
$\Omega_{crit}(q_{crit})\approx \omega_+(q_{crit})$ there are no
further collective excitations.
The related damping rates are shown in Fig. \ref{dispersionT0_paper} {\bf (B)}.
We observe that above a critical wave vector we cannot find 
collective excitations which represents a pure quantum effect. 
This critical wave vector $q_{crit}$ is determined by the used 
interaction $V_0$. Now we can link
the minimal interaction required for collective oscillation with the mass
number by equation (\ref{scalingL}). The relation between 
minimal interaction $V_{0_{crit}}$ and 
mass number $A_{crit}$ (wave vector $q_{crit}$) is plotted 
in Fig. \ref{AkritvsV0krit_paper}.
As a result we find the following fit
\be
V_{0_{crit}}(A_{crit})=(a+\frac{b}{A_{crit}}) \,\,\mbox{MeV fm}^3\, ,
\label{scale2}
\ee
where $a=151.9$ and $b=2997.4$.
The marked area Fig. \ref{AkritvsV0krit_paper} designates a region
where no collective excitations exist.
One sees that a certain interaction strength is necessary in order
to have collective oscillations. The physical meaning is that for light nuclei
with given interaction strength the sound velocity couples to the single
particle motion more strongly. This fact is reflected in the critical 
wave vector, where the collective mode enters the pair 
continuum (cp. Fig. \ref{dispersionT0_paper}).

%--------------------------Fig. 4 --------------------------------------------

Since this model relies on the simple scaling law (\ref{scalingL}) 
the equation (\ref{scale2}) can only be a qualitative estimation.
For finite nuclei we expect different parameters $a$ and $b$.

\begin{figure}
\centerline{\psfig{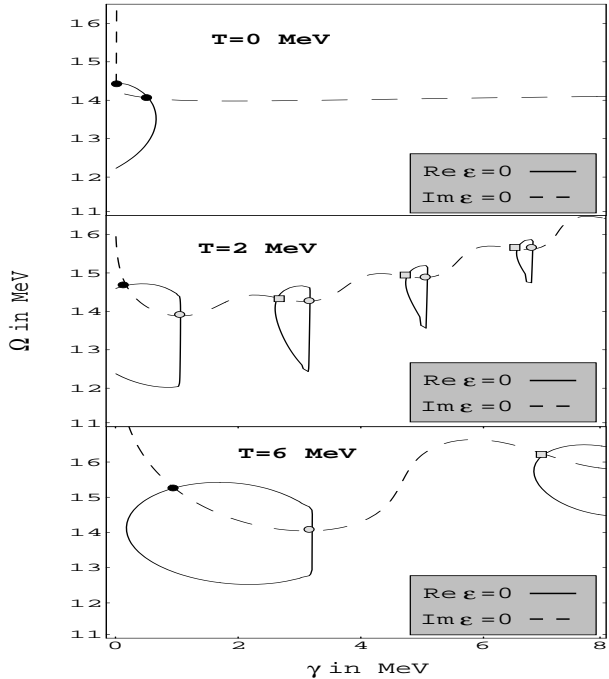}}
\vspace{.5cm}
 \caption{Collective excitations (thick dots) corresponding to the zeros of
          the analytic continuation of the complex DF (Lindhard)
          at different temperatures compared with the $T=0$ result (top or
          Fig. \ref{crossT0_paper} (top)).  }
                      \label{crossT_bild}
\end{figure}

\begin{figure}
\centerline{\psfig{file=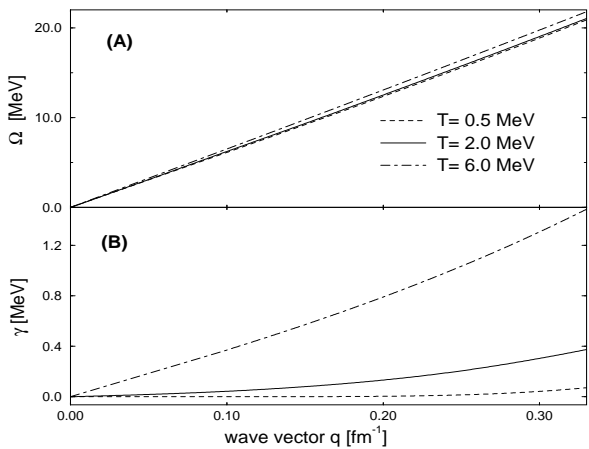,height=6.1cm,width=8cm,angle=-90}}
\vspace{.5cm}
 \caption{ Collective excitations at $T=2$ MeV (solid line) and $T=6$ MeV
  (dot-dashed line) of Fig. \ref{crossT_bild}. {\bf (A)} Dispersion of the
  collective excitations, {\bf (B)} damping rates. }
                      \label{dispersionT_paper}
\end{figure}

\begin{figure}
\centerline{\psfig{file=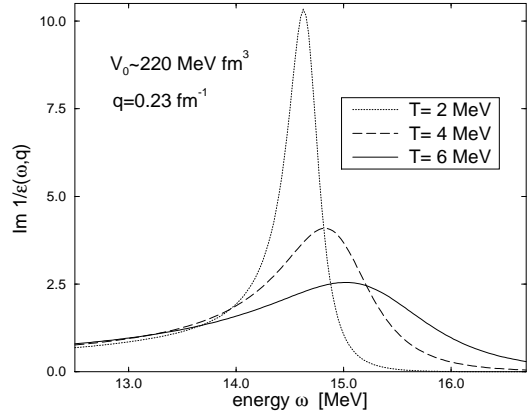,height=6.5cm,width=8cm,angle=-90}}
\vspace{.5cm}
 \caption{Imaginary part of the inverse Lindhard DF (spectral function)
          for different temperatures and the wave vector $q=0.23$ fm$^{-1}$.
          (This figure corresponds to the results of \protect\cite{BRV94}). }
                      \label{inversEps_temp_pot_paper}
\end{figure}

Considering now nuclear matter for finite temperature we present in
Fig. \ref{crossT_bild} the zeros of the real and imaginary part of the DF
compared with the $T=0$ result.
We find with growing temperatures
a continuous deformation of the lines Re$(\epsilon^c)=0$ and
Im$(\epsilon^c)=0$. The crossing points (black marked) correspond
again to collective excitations and are always damped. 
We have additional zeros (light marked) in
contrast to the zero temperature result. They
correspond to the complex poles of equation (\ref{matsubara}) and
can be identified as single
particle excitations. Taking the entire
dispersion relation for different temperatures (Fig. \ref{dispersionT_paper})
we observe higher damping rates of the
collective excitations as the temperature increases.
This effect of temperature is also visible in the  calculated spectral function
(\ref{inverseps}) of Fig. \ref{inversEps_temp_pot_paper}.
%--------------------------Fig. 5 --------------------------------------------

%--------------------------Fig. 6 --------------------------------------------
Growing temperatures lead to
a broadening of the peak and reaching a temperature of 6 MeV (solid line
Fig. \ref{inversEps_temp_pot_paper}) only small
collective effects remain observable.
The corresponding centroid energy is increasing with higher temperatures.
The further inclusion of correlations will lead to a decreasing of the
centroid energy for higher  temperatures, which has been demonstrated
in \cite{KLT97,BRV94}.
%--------------------------Fig. 7 --------------------------------------------

In Fig. \ref{damping_GDR_paperTheo} we see now for the two nuclei
$^{120}$Sn and $^{208}$Pb (dotted lines) the calculated
(Landau) damping rates of GDR over the
nuclear temperature. These damping rates are compared and discussed with other
approximation in the next sections.

\subsection{BUU equation-relaxation time approximation}
\label{BUU}
In order to consider collisions as an additional damping effect of GDR
we start first from (\ref{kinpoisson}) with the Markovian
Boltzmann  (BUU) collision integral (\ref{boltzmannstoss1}) $I_B$.

Within the standard treatment \cite{LPI79} we linearize
the collision integral around a homogeneous
equilibrium distribution $f^0$
\be
\delta f_i(p,R,T)&=&f_i(p,R,T)-f_i^0(\epsilon_p)\nonumber\\
&\equiv&\Phi_i(p,R,T) \frac{\partial}{\partial\epsilon_p}f_i^0
                            \label{ansatzlinear1}
\ee
where 
$f_i^0$ should be first the global Fermi distribution.
Later on we will use a local equilibrium distribution in order to ensure
conservation laws. The linearized collision integral reads
\be
I_B&=&\frac{1}{T}\int \frac{d^3p_2 d^3p_3} {(2\pi)^6} \,W(p_1p_2,p_3p_4)
\bigtriangleup\!\Phi\,\delta(\bigtriangleup\epsilon)\nonumber\\
&\times&[f_1^0 f_2^0(1-f_3^0)(1-f_4^0)] \, ,
\label{boltzmannstoss1linear}
\ee
where
$\bigtriangleup\epsilon=\epsilon_1+\epsilon_2-\epsilon_3-\epsilon_4$ and
$\bigtriangleup\Phi=\Phi_1+\Phi_2-\Phi_3-\Phi_4$.
Neglecting the backscattering terms $\Phi_2-\Phi_3-\Phi_4$ we obtain
the relaxation time approximation
\be
I_B=-\frac{f-f^0}{\tau(p_1)}=-\frac{\delta f}{\tau(p_1)}\, ,
     \label{ansatztau}
\ee
with
\be
\frac{1}{\tau_B(p_1)}&=&\int\frac{d^3p_2 d^3p_3}{(2\pi)^6}\,
W(12,34)\,\delta(\bigtriangleup\epsilon)\nonumber\\
&\times&\frac{f_2^0(1-f_3^0)(1-f_4^0)}{1-f_1^0}.
\ee
Furthermore we will use a thermal averaged quantity. 
This procedure is introduced in appendix \ref{thermal} 
and we get from (\ref{ansatztau})
\be
\frac{1}{\tau_B}&=&\frac{1}{n_D}
\int\frac{d^3p_1 d^3p_2 d^3p_3}{(2\pi)^9}\,
W(p_1p_2,p_3p_4)\,\delta(\bigtriangleup\epsilon) \nonumber \\
&\times&[f_1^0f_2^0(1-f_3^0)(1-f_4^0)]\, ,\label{tauboltzmann}
\ee
with $n_D=mp_FT/(2\pi^2)$.
For temperatures which are small compared to the Fermi energy we
follow the well known methods of the Fermi liquid theory \cite{SMJ89,BPE91}
and get from (\ref{tauboltzmann})
\be
\frac{1}{\tau_B}=\frac{2 \pi m^4p_f}{n_D}
\left<{W(\theta,\phi)\over \cos{\theta/2}}\right>\,I^f_B\,\label{fermiB}
\ee
where
\be
\big<...\big>=\frac{1}{(2\pi)^9}
\int\limits_0^{\pi}\,d\theta\sin\theta
\int\limits_0^{2\pi}\,d\phi
\int\limits_0^{2\pi}\,d\phi_2\,\frac{W(\theta,\phi)}{\cos{\theta/2}}\, .
                               \label{tauboltzmannana1}
\ee
The calculation of the Fermi integral $I^f_B$ is performed
in appendix \ref{integral} with the result
\be
I^f_B=\frac{2\pi^2}{3}T^3\, .
\ee
Using the collision probability in Born approximation we introduce the
spin-isospin averaged cross section $\frac{d\sigma}{d\Omega}$
\be
W(\theta,\phi)=2\pi\left |\frac{4\pi}{m}\right|^2\frac{d\sigma}{d\Omega}\,\,g
\ee
with the spin-isospin degeneracy $g=4$.
Assuming a mean cross section
$\frac{d\sigma}{d\Omega}\approx \frac{\sigma}{4\pi}$ we get finally
\be
  \frac{1}{\tau_B}=\frac{8}{3}T^2 m\,\sigma  \, ,
                                      \label{tauboltzmannana}
\ee
where $\sigma$ is the nuclear cross section (for numerical calculation
we use $\sigma=40$ mb as a spin-isospin averaged cross section
of $\sigma_{nn}$, $\sigma_{pp}$ and $\sigma_{np}$ \cite{ABO92}).

%--------------------------Fig. 8.1-------------------------------------------

\begin{figure}
\centerline{\psfig{file=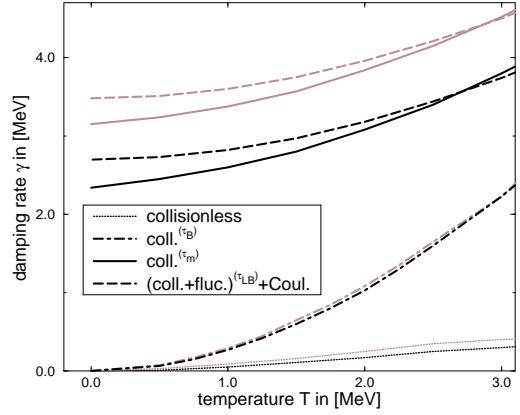,height=6.5cm,width=8cm,angle=-90}}
\vspace{.5cm}
 \caption{Theoretical damping rates of GDR for $^{120}$Sn (grey) and 
          $^{208}$Pb (black) as a function
          of the nuclear temperature $T$. They are the complex solution 
          of the dispersion relation for different approximations: 
          collisionless (Lindhard) DF (dotted line),
          Mermin DF (Boltzmann) (dot-dashed line), Mermin DF (memory)
          (solid line) and Mermin DF (density fluctuation+Coulomb
          interaction) (dashed line). }
                                         \label{damping_GDR_paperTheo}
\end{figure}
Based on this result one can use the relaxation time approximation
to include collision effects into the DF.
This leads to a pure replacement $\omega\rightarrow\omega+i/\tau$ in
the collisionless Lindhard DF (\ref{Lindhard_polarisation})
and is known to violate the sum rule (\ref{f_summenregel}).
Therefore we apply an extended approach using a conserving
relaxation time approximation following Mermin \cite{MER70}.
The local equilibrium distribution $f^0$ is characterized by
\be
f^0=\frac{1}{\exp[\beta(\epsilon_p-\mu-\delta \mu(R,t))]+1}\, ,
\ee
where the local chemical potential $\delta\mu$ is determined by
the density conservation resulting in
$\delta\mu(q,\omega)=\frac{\delta n(q,\omega)}{\Pi^0(q,0)}$. The
linearized kinetic equation (\ref{kinpoisson_linear}) reads then
\be
\lefteqn{\left(\omega-\frac{q\cdot p}{m}\right)\delta f(p,q,w)=}\hspace{1cm}
\nonumber\\
&&\left[f^0(p-q/2)-f^0(p+q/2)\right]\,\delta V(q,\omega) \nonumber \\
&&-\frac{1}{\tau_{B}}\left(\delta f(p,q,\omega)+
\frac{\partial f^0(p)}{\partial\epsilon_p}
\frac{\delta n(q,\omega)}{\Pi_0(q,0)} \right)
\ee
and the resulting (Mermin) DF is
\be
\epsilon^M(q,\omega)=1+\frac{\displaystyle \left(1+\frac{i}{\omega\tau_{B}}
\right)
\left[\epsilon(q,\omega+i/\tau_{B})-1 \right] }
{\displaystyle 1+\left(\frac{i}{\omega\tau_{B}}\right)
\frac{\left[\epsilon(q,\omega+i/\tau_{B})-1 \right]}
{\left[\epsilon(q,0)-1 \right]} }\, . \label{merminDF}
\ee
Here $\epsilon$ is the collisionless Lindhard DF (to calculate the
zeros of $\epsilon^M$ one has to use the analytical continuation of
Lindhard DF equation (\ref{landauDF}))
and $\tau_B$ the relaxation time (\ref{tauboltzmannana}).
Further conservation laws can be incorporated \cite{HPR93}.

\begin{figure}
\centerline{\psfig{file=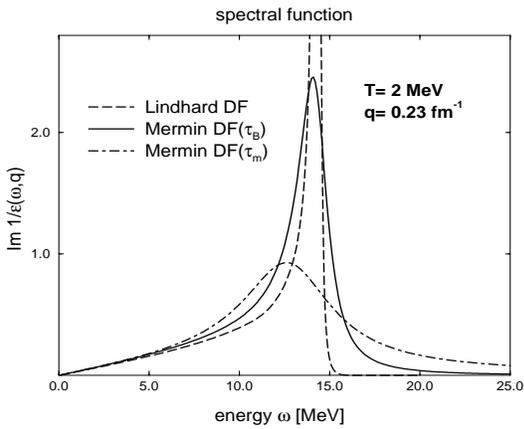,height=6cm,width=8cm,angle=-90}}
\vspace{.5cm}
 \caption{Spectral function (\ref{inverseps}) with 
          the Lindhard DF (dashed line) compared with the
          Mermin DF for different relaxation times: Boltzmann
          relaxation time $\tau_B$ (solid line), 
          dynamical (memory) dependent relaxation time $\tau_m$
          (dot-dashed line).}
                     \label{merminDF_paper}
\end{figure}

In Fig. \ref{merminDF_paper} we compare the spectral functions
computed with the Lindhard DF (dashed line) and Mermin DF (solid line), 
respectively.
We observe an additional broadening beyond Landau damping
in the width of the Mermin DF due to collision damping in this approximation.
In comparison with Fig. \ref{inversEps_temp_pot_paper} we see further
that the centroid energy of Mermin result (solid line) 
is shifted to lower values with increasing
temperature, while the centroid energy of the Lindhard result (dashed line) 
is shifted to higher values.
The result for the damping rates $\gamma$ of the GDR of
$^{120}$Sn and $^{208}$Pb is shown in Fig. \ref{damping_GDR_paperTheo}
(dot-dashed lines) which are determined by the zeros of the Mermin DF
$\epsilon^M(\Omega, \gamma)=0$.
The calculated damping rates in this approximation present  
an improvement compared to the collisionless results (dotted lines). 
We observe the typical $T^2$ behavior of the
damping known from experimental data (Fig. \ref{damping_GDR_paperFWHMexp}) 
but we do not get any contribution at zero temperature.
Now we shall employ a better approximation of collisions in the next section.
%%--------------------------Fig. 9 ----------------------------------------

%***********Unterkapitel dynamische Relaxationszeit************************

\subsection{Non-Markovian equation-dynamical relaxation time}
\label{nonMarkov}

Next we investigate the kinetic equation with the collision
integral (\ref{memorystoss1}) taking into account memory
effects.
The linearization of the collision term $I_m$ (\ref{memorystoss1})
can be performed in the same way as above with an additional
linearization with respect to $U^\pm$. Expanding $U^\pm$
in (\ref{memorystoss1}) gives
\be
\delta U^\pm(T-\tau,T)=U_0^\mp(\tau)
\big[1\pm i \int\limits_T^{T-\tau}dt^\prime\,\bigtriangleup\delta\epsilon
\big] \, ,\label{zeitoperator3}
\ee where $\bigtriangleup\delta\epsilon=\bigtriangleup\epsilon(t^\prime)-
\bigtriangleup\frac{p^2}{2m}\,\, \mbox{and}\,\,
U_0^\pm(\tau)=\exp(\pm i(\bigtriangleup\frac{p^2}{2m}\tau))$.
The linearization of
the collisional term (\ref{memorystoss1}) reads now
\be
I_{m}(p_1,T)&=&\int\limits_{0}^{\infty}d\tau\int\frac{d^3p_2 d^3p_3} {(2\pi)^6}
\,W(12,34)\nonumber \\
&\times& \bigg[ U_0^+(\tau)+U_0^-(\tau)(\delta F_1-\delta F_2)
\nonumber \\
&&\bigg.+\delta U^+(T-\tau,T)-\delta U^-(T-\tau,T)\,F^0 \bigg]
                                             \label{memorystoss1linear}
\ee
where
\be
F^0=
&&[f_3^0 f_4^0(1-f_1^0)(1-f_2^0)-f_1^0 f_2^0(1-f_3^0)(1-f_4^0)].\nonumber\\
&&
\ee
The calculation of $\delta F_1$ with the help of
(\ref{ansatzlinear1}) results into
\be
\delta F_1&=&f_3^0 f_4^0(1-f_1^0)(1-f_2^0)\nonumber\\
&\times&\frac{[f_1^0\Phi_1+f_2^0\Phi_2-(1-f_3^0)\Phi_3-(1-f_4^0)\Phi_4 ]}{T}
                                                            \label{f1_linear}
\ee
and $\delta F_2$ is obtained from equation (\ref{f1_linear}) by
interchanging $f\leftrightarrow (1-f)$.
According to equation (\ref{kinpoisson_linear}) we perform a
Fourier transformation  $T\rightarrow\omega$ of the linearized collision
integral (\ref{memorystoss1linear}) with the result
\be
I_{m}(p_1,\omega)&=&\frac{1}{2}\int \frac{d^3p_2 d^3p_3} {(2\pi)^6}\,W(12,34)
                                                               \nonumber\\
&\times&\left [D_+(\omega)\,\left(\delta F_1-\delta F_2\right)+
\frac{D_-(\omega)}{\omega}\bigtriangleup\delta\epsilon\right] \,,\label{endmem}
\ee
where $D_\pm(\omega)=\delta(\omega+\bigtriangleup\epsilon)\pm
\delta(\omega-\bigtriangleup\epsilon)$. 
For $\delta\epsilon$ in (\ref{endmem}) we obtain
from the dispersion relation (\ref{e_dispersion}) 
\be
\delta\epsilon_(p_i,R,T)&=&\frac{1}{1-\frac{\partial Re\Sigma(p_i,R,T)}
{\partial\Omega}|_{\Omega=\epsilon}}\Big[
U^\prime(n_0)\delta n(R,T) \Big.\nonumber\\
&+&\Big.\sum_{\bar{p}}\
\frac{\partial\Sigma(p_i,R,T)}{\partial f(\bar{p},R,T)}\delta f(\bar{p},R,T)
\Big]\, , \label{deltaeps}
\ee
where $\delta n(R,T)=\sum_{\bar p}\delta f(\bar{p},R,T)$. In the
following we neglect self-energy effects ($\Sigma=0$) and restrict our
consideration to quasi-particle energies in mean-field approximation
$\delta\epsilon(p_i,R,T)=U^\prime(n_0)\delta n(R,T)$ with
$U^\prime=V_0$ from equation (\ref{Potential_V0}). Therefore we see that
$\bigtriangleup\delta\epsilon=0$. This would be not the case for
momentum dependent potentials.

One can now introduce a thermal averaged relaxation time $\tau_{m}$
analogous to (\ref{tauboltzmann})
\be
\frac{1}{\tau_m(\omega)}&=&-\frac{1}{2n_D}\int\frac{d^3p_1 d^3p_2 d^3p_3}
{(2\pi)^9}\,W(12,34) \nonumber \\
&\times&\bigg\{D_+(\omega)\,f_1^0\bar{f_1^0}
\left(\bar{f_2^0}f_3^0f_4^0-f_2^0\bar{f_3^0}\bar{f_4^0}\right)\bigg\}
\, ,\label{taumemory1}
\ee
where $\bar{f}=(1-f)$ and $n_D$ is specified in appendix \ref{thermal}.
Comparing the Markovian result equation (\ref{tauboltzmann})
with (\ref{taumemory1}) we observe a replacement of the
energy conserving factor $\delta(\epsilon)$
with $\delta(\omega+\bigtriangleup\epsilon)$ and
$\delta(\omega-\bigtriangleup\epsilon)$. The energy $\hbar\omega$ of the
collective excitations are now included into the energy conservation.
We can interpret this effect as a coupling of the collective modes to
the binary collisions. The collective boson is absorbed or emitted
when two particles are colliding \cite{ABO92}.
Restricting the relaxation time $\tau_{m}$ again to temperatures which
are small compared to the Fermi energy we follow (\ref{tauboltzmann})
and get from (\ref{taumemory1})
\be
\frac{1}{\tau_m(\omega)}=\frac{2 \pi m^4p_f}{n_D}
\left<{W(\theta,\phi)\over \cos{\theta/2}}\right>\,I^f_m(\omega)\,
\label{fermiM}
\ee
where the Fermi integral $I^f_m$ is done in appendix \ref{integral} with the
result
\be
I^f_m(\omega)=T^3\left[\frac{2}{3}\pi^2+
   \frac{1}{2}\left(\frac{\omega}{T}\right)^2\right].
\ee
The analytical expression for $\tau_{m}$ finally reads
\be
 \frac{1}{\tau_{m}(\omega)}=\frac{1}{\tau_B}\left[1+
 \frac{3}{4}\left(\frac{\omega}{\pi T} \right)^2 \right] \, ,\label{taumemory2}
\ee
where $\tau_B$ is the relaxation time (\ref{tauboltzmannana})
which corresponds to the Markovian limit.
The expression (\ref{taumemory2}) 
for the relaxation time $\tau_m$ contains
an additional (dynamical) contribution ($\sim \omega^2$).
It guarantees that also at zero temperatures the collective mode can be
considered as a self-propagating one which has a finite damping.
We obtain a relaxation rate which is the Landau damping rate of zero 
sound \cite{LPI79} except the factor
of 3 in front of the frequencies. 

The form (\ref{taumemory2}) is similar to the result Ayik
derived in \cite{ABO92} for the collective damping
rate. In contrast to this result we have here the single-particle relaxation time.
As long as we have only momentum
independent mean fields, the linearization of the mean field propagator
does not reduce this to the standard Landau result as claimed in
the errata of \cite{ABO92}, instead the factor of 3 counts here
(see comment after equation (\ref{deltaeps})).

One may argue that this factor 3 comes from the thermal averaging 
we employed here. We can repeat all calculations without 
thermal averaging (cp. appendix \ref{nointegral})
with the momentum in the relaxation time at the Fermi level. 
The resulting dynamical relaxation time reads then
\be
 \frac{1}{{\tilde\tau}_{m}(\omega)}=\frac{3}{4}\frac{1}{\tau_B}\left[1+
  \left(\frac{\omega}{\pi T} \right)^2 \right] \, .\label{notaumemory}
\ee
Compared with (\ref{taumemory2}) one sees no factor $3$ appears in front of the
frequency, however a factor $3/4$ in front of the relaxation time $1/\tau_B$.
Therefore we obtain for $T=0$ the same frequency dependence in both cases. In other
words the zero sound damping is not affected by averaging. 
The temperature dependence is slightly flatter without thermal averaging. 

In the Mermin approximation $\epsilon^M $ of (\ref{merminDF}) 
we have to replace the
static relaxation time $\tau_B$ with the dynamical one $\tau_m(\omega)$.
Then we can again compute the damping rates $\gamma$ for the GDR of
$^{120}$Sn and $^{208}$Pb
by searching for the zeros of the
Mermin DF $\epsilon^M(\Omega,\gamma)=0$.
The results are shown in Fig. \ref{damping_GDR_paperTheo} (solid lines). 
We observe a $T^2-$ dependence of the damping which is
due to the low temperature expansion of the relaxation time $\tau_m$,
but the temperature increase is too flat compared with the Boltzmann 
result (dot-dashed line).
The inclusion of memory effects now induces a finite width at
low temperatures. This reflects the fact that memory accounts for zero 
sound damping. 
Next we investigate the influence of density oscillations 
on the potential itself.

%************Unterkapitel: Dichtefluktuationen***************************

\subsection{Contribution of density fluctuations}
\label{fluk}

As shown in \cite{MOT96} the contribution of density fluctuations can have
an remarkable effect on the damping rate. The
derivation of the kinetic equation including the density 
oscillations leads to Lenard-Balescu type of collision integrals \cite{MOT96}
which have the same structure as (\ref{memorystoss1}) but with a dynamical
potential
\be
V^2(q)\rightarrow {V^2(q) \over |\epsilon(q,\epsilon_1-\epsilon_1')|^2}\, ,
\label{dynamiPot}
\ee
replacing the static one. Here the DF renormalizes the potential. Then we can
proceed as described above and linearize the
collision integral, etc. This would lead to a rather involved integration.
We simplify the treatment by assuming that the
transfered momentum during a single collision is small. 
Since for small $q$ we have for the Lindhard DF
\be
\lim\limits_{q\rightarrow0} \epsilon(q,\omega)=
             1-\frac{c^2q^2}{\omega^2}+{\cal O}(q^4)
\ee
with the free sound velocity $c^2=V_0 n/m$, we get from (\ref{merminDF})
\be
&&\lim\limits_{q\rightarrow0} \epsilon^M(q,\omega)=
1-\frac{c^2 q^2}{\omega(\omega+i/\tau)}\, . \label{lin}
\ee
Inspecting (\ref{taumemory1}) we see that the frequency argument of
(\ref{dynamiPot}) has to be taken at
$\epsilon_1-\epsilon_1'= q {2 p_2+q \over 2m}\mp \omega\approx\mp \omega$
if we fix the final state of scattering as the
state which matters for kinetic processes and use vanishing transfer
momentum $q$. Further we observe that
$|\epsilon(q,\omega)|=|\epsilon(q,-\omega)|$ such that we have a common
factor in (\ref{taumemory1}) in front of $D_+$ of
$1/|\epsilon(q \rightarrow 0,\omega)|^2$. The further procedure is as above
described. We expand for low temperatures and
observe that for the angular averaging holds
\be
\left<\frac{q^2}{\cos{\theta/2}}\right>=                              
\left<\frac{2p_F\sin(\theta/2)\sin(\phi/2)}{\cos{\theta/2}}\right>\,. 
\ee
Using (\ref{lin}) the resulting prefactor which
renormalizes (\ref{taumemory1}) reads 
\be
{1 \over \tau_{LB}(\omega)}&=&{1 \over \tau_m(\omega)}
\int\limits_0^1dx\,\left|\epsilon^M(2p_Fx,\omega)\right|^{-2}\nonumber\\
&=&{1 \over \tau_m(\omega)}\int\limits_0^1dx\frac{1}{1-2 {\rm Re}
z x^2+|z|^2 x^4}
\label{LB_tauI}
\ee
with
\be
z={4 c^2 p_f^2\over \omega (\omega +{i\over \tau})}
\ee
and the complex frequency $\omega=\Omega-i \gamma$.
With tabulated integrals \cite{GRY94} we get from (\ref{LB_tauI})
finally an analytical expression
\be
{1 \over \tau_{LB}(\Omega,\gamma)}
&=&{1 \over \tau_m(\Omega,\gamma)}\frac{k}{2\sin 2\alpha}\nonumber\\
&\times&\bigg[\sin \alpha\,{\rm arctanh}(2k\cos{\alpha}+k^2)\nonumber\\
&&+\cos{\alpha}\Big(
\arctan{\big(\frac{1-k^2}{2k\sin{\alpha}}\big)}+\frac{\pi}{2}\Big) \bigg]\, ,
\label{tauLB}
\ee
where $\alpha=\frac{1}{2}\arccos{({\rm Re}z/|z|^2)}$ and $k=1/\sqrt{|z|}$.
For our chosen situation and potential
the prefactor (\ref{tauLB}) increases the damping rates only by $0.2$ MeV. 
However, we see that the renormalization of the potential 
by density fluctuations which
are in turn determined by the linearization of
the kinetic equation can lead in principal to a remarkable 
change in the dispersion relation if we come close to 
the instability line \cite{MOT96}.
Our model potential (\ref{Potential_V0}) has no instability in the
parameter range.
The procedure here means that density fluctuations
are caused by interactions or correlations,
however these density fluctuations renormalize the potential, i.e. their
cause. So we have a complicated feedback of
correlations to the fluctuations and so forth.

\subsubsection{Inclusion of Coulomb effects}

We expect a more pronounced effect if long range interactions mediate
collective oscillations.
The Coulomb interaction leads to an
additional contribution for the proton meanfield (\ref{PotNeu}) 
\be
V^C=\frac{4 \pi e^2}{q^2}n_p(q)\,,
\ee
so that the  resulting dispersion relation will be a 
matrix equation \cite{MFW97}. As above, the damping rates $\gamma$ 
are again the complex solutions of this dispersion relation and plotted 
in Fig. \ref{damping_GDR_paperTheo} (dashed lines). For small temperatures
we find an increasing of the damping rates compared 
with the rates of the memory
dependent collision approximation (\ref{taumemory2}) (solid lines).
This comes from the fact that we have density fluctuations at zero temperature
caused by the Coulomb interaction.
For temperatures larger than $3$ MeV the binary collisions dominate and
the behavior follows the memory collision approximation.

\subsection{Comparison with the Experiment}
\label{exper}

It is interesting to point out an advantage of the 
Mermin polarization function.
Therefore we compute the power spectrum of the mode, i.e. 
the energy rate per time which is expended 
on the motion of the collective mode. This power spectrum is connected to 
the structure factor 
$S(q,\omega)=\sum\limits_f |<f|V_0|0>|^2 \delta(\hbar \omega -E_f +E_0)$
by
\be
P(q,\omega)= {2 \pi \over \hbar} S(q,\omega) \hbar \omega
\ee
which is just Fermi's Golden Rule. The structure factor itself is given by 
the dielectric function 
$S=V_0/\pi {\rm Im} \epsilon^{-1}$. 
Using (\ref{lin}) we arrive at an expression for the power 
spectrum of
\be
P(q,\omega)&=&2 V_0\omega \; {\rm Im} {\omega (\omega +
{i\over \tau})\over \omega (\omega +{i\over \tau})-
{V_0n q^2 \over m}}\nonumber\\
&=&2 V_0 \omega_0^2 {{\omega^2\over \tau}\over (\omega^2-\omega_0^2)^2+
({\omega\over \tau})^2} \label{lorentz}
\ee
with $\omega_0^2=V_0n q^2/m$. 
The second line is derived with the assumption 
that $\tau$ is real, which is not 
true for dynamical relaxation times (memory effects). We rederive by this way 
just the classical Lorentz formula 
which describes the energy rate expended on the motion of a damped harmonic 
oscillator driven by the external force 
${\stackrel{..}{x}} + {\stackrel{.}{x}
}/\tau +\omega_0^2 x =2 V_0 \omega_0^2 \cos(\omega_0 t)$ averaged 
over time.
%--------------------------Fig. 10 ----------------------------------------
\begin{figure}[t]
\centerline{\psfig{file=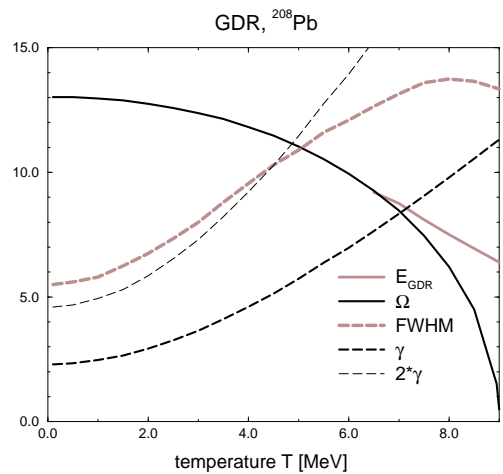,height=6.6cm,width=8cm,angle=-90}}
\vspace{.5cm}
 \caption{Temperature dependence of the centroid energy $E_{GDR}$ and of
          FWHM for GDR mode in $^{208}$Pb (grey lines) compared with
          the solution of the complex dispersion relation (\ref{complexdisp}),
          ${\rm Re}\,\omega=\Omega$ and ${\rm Im}\,\omega=\gamma$
          (dark lines), using 
          the memory dependent relaxation time (\ref{taumemory2})
          in the Mermin DF (\ref{merminDF}).}
                     \label{dampPB208Mermin_FWHM_gamma}
\end{figure}
Naturally that (\ref{lorentz}) leads to a Breit Wigner form near 
the resonance energy $\omega_0$ with the full width of 
half maximum (FWHM) of $\Gamma=1/\tau$
\be
P(q,\omega)
&=&V_0 \omega_0^2 {{\Gamma \over 2}\over (\omega-\omega_0)^2+
({\Gamma\over 2})^2}. \label{breitW}
\ee
The damping rate in classical approximation (\ref{lin}) (long wavelength)
is given as $\gamma=1/(2\tau)$.
We recognize that the FWHM is just twice the damping rate $\Gamma=2 \gamma$. 
This has been recently emphasized \cite{TKL97}.
It has to be stressed that the experimental data are accessible by FWHM.
To extract the FWHM from the structure  function (\ref{inverseps})
we used the Mermin approximation $\epsilon^M$ (\ref{merminDF}) 
with the dynamical (memory) dependent relaxation 
time $\tau_m$ of (\ref{taumemory2}).

In Fig. \ref{dampPB208Mermin_FWHM_gamma} we have plotted the temperature 
dependence  of FWHM (grey-dashed line) of the structure  
function (cp. Fig. \ref{merminDF_paper}) 
and the corresponding centroid energy $E_{GDR}$ (grey-solid line) of 
the GDR mode in $^{208}$Pb together 
with real part $\Omega$ (dark-solid line) and
imaginary part $\gamma$ (dark-dashed line)
of the complex solution of the dispersion relation, respectively.
We find an approximate relation $FWHM\approx 2\gamma$  
(grey-dashed  and thin dark-dashed lines) which holds in the 
temperature limit, where $FWHM<E_{GDR}$, as has been 
discussed in \cite{TKL97}.

We emphasize that the centroid energy $E_{GDR}$ of the structure function 
agrees with the
real part $\Omega$ of the complex solution of dispersion relation up to higher
temperatures (thick-dark and thick-grey lines).
The small deviation from the exact Breit Wigner form (\ref{breitW})   
is caused by the memory effects resulting in
a frequency dependent dynamical relaxation time $\tau_m$ which is now complex.
%--------------------------Fig. 8.2-------------------------------------------
\begin{figure}
\centerline{\psfig{file=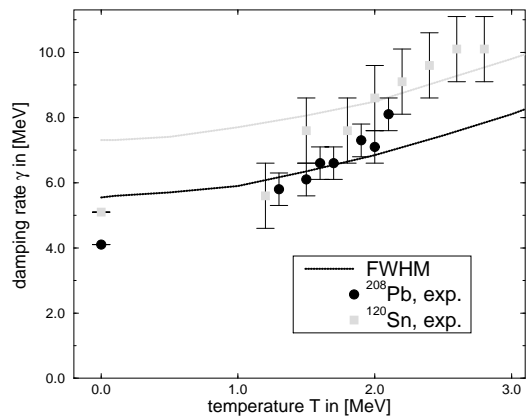,height=6.5cm,width=8cm,angle=-90}}
\vspace{.5cm}
 \caption{Experimental damping rates of GDR for $^{120}$Sn and $^{208}$Pb 
           ($^{120}$Sn from \protect\cite{RAP96} and $^{208}$Pb from
          \protect\cite{RAM96}) compared with the FWHM of the structure 
          function as a function of the nuclear temperature $T$.} 
                                            \label{damping_GDR_paperFWHMexp}
\end{figure} 
The FWHM is the observable which allows us to compare with the 
experimental data.
In Fig. \ref{damping_GDR_paperFWHMexp} we have plotted 
the FWHM of the GDR modes in $^{120}$Sn (grey line) and 
$^{208}$Pb (dark line) as a function of temperature 
compared with experimental data. 
They coincide within theoretical limits if we consider 
that the experimental values are fits of a Breit Wigner form itself.

\begin{figure}
\centerline{\psfig{file=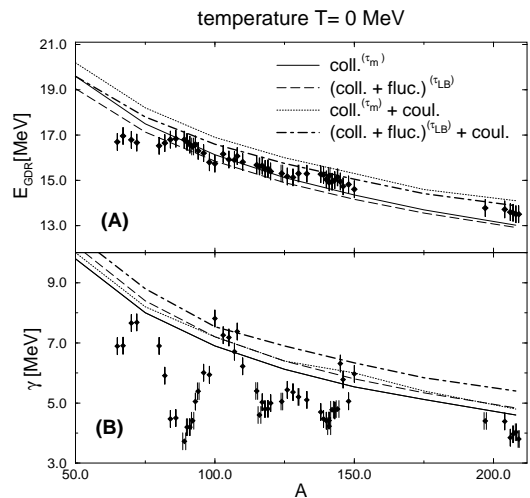,height=6.5cm,width=8cm,angle=-90}}
\vspace{.5cm}
 \caption{The experimental centroid energy $E_{GDR}$  {\bf(A)} and 
          the experimental data of the damping widths       
          $\gamma$ {\bf(B)} for GDR \protect\cite{BER88} 
          vs mass number          
          together with different approximations.}
                  \label{ga_A_T0}
\end{figure}

In Fig. \ref{ga_A_T0} we compare different approximations calculated in 
this  paper with experimental data versus mass number for $T=0$. 
In {\bf (A)} we have plotted the centroid energy $\Omega=E_{GDR}$.
We observe that the successive inclusion of collisions with the memory 
dependent relaxation time $\tau_m$ (\ref{taumemory2}) (thin solid line), 
density fluctuations with the relaxation time 
$\tau_{LB}$ (\ref{tauLB}) (dashed line) as well as the inclusion of the
Coulomb interaction (dotted and dot-dashed line) reproduces the experimental 
values (diamonds) increasingly for the mass numbers $80\leq\mbox{A}\leq 210$.  
The inclusion of only Coulomb effects (dotted line) slightly overestimates 
the data. This overestimate is compensated if we add the 
density fluctuations (dot-dashed line).
The corresponding FWHM, which are shown to be roughly twice the 
imaginary parts $\gamma$ of the complex dispersion relations are plotted 
in Fig. \ref{ga_A_T0}{\bf (B)}.
The inclusion of collisions brings the curve towards the experimental values.
The improvement by inclusion of Coulomb effects and density fluctuations is
small. 
            
%--------------------------Fig. 11 ----------------------------------------

%************Unterkapitel: Zusammenfassung*********************************

\section{Summary}
\label{sum}
We have presented a systematic microscopic description of collective
excitations in hot nuclear matter by a kinetic approach and computed
damping rates of giant dipole resonances (GDR).
We have rederived a generalized quantum kinetic equation
employing the real-time Green functions technique.
Within Hartree approximation one derives the
collisionless Vlasov equation which in the linearized limit leads
to the known Lindhard DF in RPA. Applying the complex integration method
proposed by Landau we calculated
the damping width (Landau damping) of GDR in different nuclei
at finite temperatures. 

The experimental data are calculated via the FWHM.
Landau damping rates 
underestimate the experimental data of $^{120}$Sn und $^{208}$Pb.
Considering collisions in
first Born approximation the (Markovian) kinetic equation is of
Boltzmann or BUU type. Using a relaxation time approximation
we have incorporated the collision effects into
the DF which leads to the Mermin DF. The calculated damping rates of the
GDR show a           
typical $T^2$-behavior but do not reproduce the experimental values yet.
Only memory effects which are included into the collision term
of the kinetic equation via retardation of the distribution function
improve the theoretical damping rates. This 
non-Markovian relaxation time approximation leads to a
dynamical relaxation time which differs from the
Landau result of the damping of zero sound. Within the
low temperature limit the theoretical damping rates of the GDR
in $^{120}$Sn und $^{208}$Pb are improved but
the temperature increase is flatter than the experimental findings.
A renormalization of the potential by density fluctuations increase the 
damping rates by $\approx 7 \%$. These density fluctuations 
together with Coulomb effects reproduce the centroid energy for the mass
numbers $80\leq \mbox{A}\leq 210$. If we neglect Coulomb interaction 
we underestimate the centroid energy for heavier elements.
                     
We consider the solution of complex dispersion relations as the physical mode
of collective excitation in the system. In contrast, the experimentally
observed collective excitations are extracted from FWHM.
The calculation of the width (FWHM) of the GDR modes,
which is proportional to
imaginary part of the inverse DF, leads to results which are 
comparable to the experiment. Memory effects change the shape
of these structure functions compared to the usual Breit Wigner form.
Since experimental data are fitted with the latter one, we cannot expect
to reproduce the experimental data completely. Especially the temperature
dependence of $\gamma$ as solution of the dispersion relation shows too small
increase with temperature.
This may be a hint that shape fluctuations become
important \cite{DGB96}. In \cite{DBB96} it was stated that $\gamma$
increases linearly with temperature, caused by the coupling of surface modes.
The incorporation of shape fluctuations will be considered  in
a forthcoming paper. 

\acknowledgements
The fruitful discussions with M. DiToro and V. Baran are
gratefully acknowledged.
The work was supported by the DFG (Germany) under contract Nr. Ro 905/13-1.

%--------------------Herleitungen---------------------------------------------

\appendix

\section{Generalized kinetic equation}
\label{GKEq}

We shortly sketch the derivation of the general kinetic 
equation (\ref{allgI}).
We use the dialect of the generalized Kadanoff and Baym formalism 
developed by Langreth and Wilkins \cite{LWI72} for the nonequilibrium
(real-time) Green's functions introduced by Kadanoff and Baym \cite{KBA62}.
Considering a system of Fermions which interact via the
potential $V$ we start with the Hamiltonian
\be
\hat{H}&=&\int d^3r_1\Psi_1^{\dag}\frac{\nabla^2}{2m}\Psi_1\nonumber\\
&+&\frac{1}{2}
\int d^3r_1d^3r_2\Psi_1^{\dagger}\Psi_2^{\dagger}V_{12}(r_1-r_2)\Psi_1\Psi_2
,
\ee
where the different numbers, i.e. $1=(r,t,s,...)$, denote the one-particle
variables (space, time, spin,...). The annihilation and creation
operators for Fermions obey the commutator relations
\be
[\Psi_1,\Psi_2]_+=0,\qquad[\Psi_1,\Psi_2^{\dagger}]_+=\delta_{12}.
\ee
The correlation functions are defined by different products of creation
and annihilation operators in the Heisenberg picture
\be
G^>(1,2)=\langle\Psi_1\Psi_2^{\dagger}\rangle,\qquad
G^<(1,2)=\langle\Psi_2^{\dagger}\Psi_1\rangle.         \label{korrelatioen}
\ee
Here the $\langle...\rangle$ denotes the average value with the
unknown quantum-statistical density operator $\rho$.
The causal Green function is given with the Heaviside step function
$\Theta$ by
\be
G(1,2)=\Theta(t_1-t_2)G^>(1,2)-\Theta(t_2-t_1)G^<(1,2) \label{korrelat}
\ee
It is furthermore useful to introduce the retarded and advanced
Green functions according to
\be
G^R(1,2)=&-&i\Theta(t_1-t_2)[G^>(1,2)+G^<(1,2)] \nonumber \\
G^A(1,2)=&&i\Theta(t_2-t_1)[G^>(1,2)+G^<(1,2)] \,.  \label{avan}
\ee
Using now the equation of motion of the creation and
annihilation operators we can derive kinetic equations. Applying the
equation of motion for the field operators in the Heisenberg
picture one gets an equation of motion for the causal Green
functions \cite{KBA62}
\be
\lefteqn{\left(i\frac{\partial}{\partial t_1}+\frac{\nabla_1^2}{2m}\right)
  G_1(1,1^\prime)=}\hspace{1cm}\nonumber\\
&&\delta(1-1^\prime)+\int d2\,V(1-2)G_2(12,1^\prime 2^+)\, .
                                                 \label{eom}
\ee
In this so-called Martin-Schwinger hierarchy \cite{KBA62}
the one-particle Green's function couples to the two-particle one,
the two-particle Green's function couples to the three-particle one, etc.
A formally closed equation for the one-particle equation can be reached
with the introduction of the self-energy
\be
\int\limits_c d{\bar 1}\,\Sigma(1,{\bar 1})G_1({\bar 1},1^\prime)=
\int d2\,V(1-2)G_2(12,1^\prime 2^+) \label{selbstE}
\ee
where the integration contour $c$ turns out to be chosen as Keldysh
contour in order to meet the requirement of weakening
of initial correlations
\be
&&\lim\limits_{(t_1=t_2=t_1'=t_2')\rightarrow
-\infty}G_2(12,1'2')\nonumber\\
&=&G(1,1')G(2,2')\pm
G(1,2')G(2,1')
\ee
which means that asymptotically the higher order correlations should be
factorized for infinite past.
One gets from (\ref{eom}) and (\ref{selbstE}) the Dyson equation
\be
&&\left(i\frac{\partial}{\partial t_1}+\frac{\nabla_1^2}{2m}\right)
G_1(1,1^\prime)=\delta(1-1^\prime)\nonumber\\
&+&\int d{\bar 1} \,
\left (
\Sigma(1,{\bar 1})
G_1({\bar 1},1^\prime)-
\Sigma^<(1,{\bar 1})
G_1^>({\bar 1},1^\prime)
\right ).
\label{dyson1}
\ee
Writing (\ref{dyson1}) into the more compact operator notation gives
\be
G_0^{-1}G_1=1+\Sigma G_1\, , \label{dyson2}
\ee
where $G_0^{-1}$ is the inverse of the Hartree--Fock
Green function
\be
G_0^{-1}(1,1^\prime)=
\left(i\frac{\partial}{\partial t_1}+\frac{\nabla_1^2}{2m}\right)
\delta(1-1^\prime) -\Sigma_{\rm HF}(1,1')\, . \label{gruenfrei}
\ee
We apply now the Langreth-Wilkins rules \cite{LWI72} onto the
Dyson equation (\ref{dyson2}), which describe the way to get
the correlation functions (\ref{korrelat}) and retarded functions (\ref{avan})
from causal ones. We get the equation of motion
\be
\left(G_0^{-1}-\Sigma^R\right)G^{\stackrel{>}{<}}=
                                 \Sigma^{\stackrel{>}{<}}G^A\, . \label{d1}
\ee
Subtracting the adjoined equation the general
kinetic equation reads \cite{SLI95}
\be
\lefteqn{ -i\left(G_0^{-1}G^<-G^<G_0^{-1}\right)=}\hspace{1cm}\nonumber\\
&&         i\left(G^R\Sigma^<-\Sigma^<G^A\right)-
           i\left(\Sigma^RG^<-G^<\Sigma^A\right) \, ,\label{akbE}
\ee
which was derived first by Kadanoff and Baym \cite{KBA62}.
For the time diagonal case $t_{1}=t_{1^\prime}=t$ and using
(\ref{avan}) in (\ref{akbE}) we finally get
\be
&&-i\left[G_0^{-1}(r_1t,r_{1^\prime}t),G^<(r_1t,r_{1^\prime}t)\right] =
\int d^3{\bar r_1}\int\limits_{-\infty}^{t}d{\bar t_1}          \nonumber\\
\bigg(&&G^>(r_1t,{\bar 1})\Sigma^<({\bar 1},r_{1^\prime}t)
 +\Sigma^<(r_1t,{\bar 1})G^>({\bar 1},r_{1^\prime}t)            \nonumber \\
-&&G^<(r_1t,{\bar 1})\Sigma^>({\bar 1},r_{1^\prime}t)
  -\Sigma^>(r_1t,{\bar 1})G^<({\bar 1},r_{1^\prime}t) \bigg), \label{akbE1}
\ee
which is the kinetic equation (\ref{hartreeG}) and (\ref{allgI}), respectively.

\section{Expression of the dielectric function}
\label{herleitungenA}
In this appendix we give the explicit expressions of the
real and imaginary parts of the analytical continuation of the
complex dielectric function $\epsilon^c(q,\Omega,\gamma)$
in Eq. (\ref{landauDF})
\be
{\rm Re} \,\epsilon^c(q,\Omega,\gamma)=\left\{
  \begin{array}{ll}
    \epsilon_{Re}(q,\Omega)~,                             & \gamma=0 \\
    {\rm Re} \,\epsilon_{Re}(q,\Omega-i\gamma)&\\
    - 2\,{\rm Im} \,\epsilon_{Im}(q,\Omega-i\gamma)~,         & \gamma>0 \\
    {\rm Re} \,\epsilon_{Re}(q,\Omega-i\gamma)~,              & \gamma<0 \\
  \end{array} \right.         \label{landauDFtrafore}
\ee
\be
{\rm Im} \,\epsilon^c(q,\Omega,\gamma)=\left\{
  \begin{array}{ll}
    \epsilon_{Im}(q,\Omega)~,                             & \gamma=0 \\
    {\rm Im} \,\epsilon_{Re}(q,\Omega-i\gamma)&\\
     +2\,{\rm Re} \,\epsilon_{Im}(q,\Omega-i\gamma)~,         & \gamma>0 \\
    {\rm Im} \,\epsilon_{Re}(q,\Omega-i\gamma)~               & \gamma<0~, \\
  \end{array} \right.         \label{landauDFtrafoim}
\ee
where  we redefined $\epsilon^c$  in terms of the retarded
Lindhard DF $\epsilon(q,\omega)=\epsilon_{Re}+i\epsilon_{Im}$
(\ref{Lindhard_polarisation}).
For nonzero temperature we find
\be
\epsilon_{Re}&=&1+\frac{V_0m}{\pi^2 q}
   \int\limits_0^{\infty}dk\,k\,f_F(k)\,\Big[\phi(y_+)-\phi(y_-)\Big]\, , \\
\epsilon_{Im}&=&-{V_0  Tm^2 \over \pi q}\ln\left|
                   \frac{1+\exp^{\beta(\mu-(B^\omega_-)^2/2m)}}
                        {1+\exp^{\beta(\mu-(B^\omega_+)^2/2m)}}\right|
                                                          \label{aneu}
\ee
with $y_{\pm}=B^\omega_{\pm}/k$, $B^\omega_{\pm}=\omega m/q \pm
q/2$, and
$\phi(y)=\ln\left|(1-y)/(1+y)\right|$.

Considering now complex frequency $\omega=\Omega-i\gamma$ we find for
$\epsilon_{Re}(q,\Omega,\gamma)={\rm Re} \,\epsilon_{Re}+{\rm Im} \,\epsilon_{Re}$

\be
{\rm Re} \,\epsilon_{Re}&=&
      1-\frac{V_0m}{2\pi^2 q}
\int\limits_{0}^{\infty}dk\,k\,f_F(k)\,  \label{a5}\\
&\times&\left\{\ln\frac{(k^2-(B^\Omega_+)^2-\kappa^2)^2+(2k\kappa)^2}
           {\left[(k+(B^\Omega_+)^2-\kappa^2\right]^2}\right.\nonumber\\
&&\left.-      \ln\frac{(k^2-(B^\Omega_-)^2-\kappa^2)^2+(2k\kappa)^2}
           {\left[(k+B^\Omega_-)^2-\kappa^2\right]^2}\right\}
\ee
and
\be
{\rm Im} \,\epsilon_{Re}&=&
         \frac{V_0m}{\pi^2 q}
\int\limits_{0}^{\infty}dk\,k\,f_F(k)\,  \label{a7}\\
&\times&\left[\arctan{\frac{2k\kappa}{(k^2-(B^\Omega_+)^2-\kappa^2)}}\right.
                                                                    \nonumber\\
&&\left.-\arctan{\frac{2k\kappa}{(k^2-(B^\Omega_-)^2-\kappa^2)}}\right]\, ,
\ee
where $\kappa=\gamma m/q$.\\
The expressions of ${\rm Re} \,\epsilon_{Im}$ and
${\rm Im} \,\epsilon_{Im}$ are

\be
{\rm Re} \,\epsilon_{Im}= \frac{V_0mT}{2\pi q}
                      \ln\left|\frac{\sigma_c^2+\sigma_s^2}{\sigma^2}\right|\,
\ee
and
\be
{\rm Im} \,\epsilon_{Im}= {V_0m T\over \pi q} 
\arctan\left(\frac{\sigma_s}{\sigma_c}\right)
\ee
with
\be
\sigma&=& 1+2a_-\cos(B^\Omega_+\kappa_1)+a_-^2\nonumber\\
\sigma_c&=&1+a_+\cos(B^\Omega_-\kappa_1)+a_-\cos(B^\Omega_+\kappa_1)+
            a_+a_-\cos(-\kappa_1q) \nonumber \\
\sigma_s&=&~\,\,\quad a_+\sin(B^\Omega_-\kappa_1)-a_-\sin(B^\Omega_+\kappa_1)+
            a_+a_-\sin(-\kappa_1q). \nonumber\\
&&
\ee
The coefficient $a_\pm$ is defined by
\be
a_\pm=\exp(\mu/T) \exp\left[\frac{(-(B^\Omega_\pm)^2+\kappa^2)\kappa_1}
                     {2\kappa}\right]
\ee
where $\kappa_1=\gamma/Tq$.
At zero temperature we used for the analytical continuation of the
distribution function \cite{BBS94}
\be
f_F(k)=\theta(k_F-|k|) \label{FD}
\ee
with
\be
f_F(k)&=&\lim_{\Delta\to 0}F_{\Delta}(k)~, \nonumber \\
F_{\Delta}(k)&=&\frac{1}{\pi}\left\{\arctan\left[\frac{k+k_F}{\Delta}\right]-
                             \arctan\left[\frac{k-k_F}{\Delta}\right]\right\}
\nonumber \\
\ee
The calculation of
${\rm Re}\,\epsilon_{Re}$ (\ref{a5})
and ${\rm Im}\,\epsilon_{Re}$ (\ref{a7}) for complex
frequencies will be done analogously.
The imaginary part of $\epsilon_{Im}$ in Eq. (\ref{aneu}) vanishes if
$\Delta$ goes to zero and can therefore be dropped.

%--------------------------------------------------------------

\section{The thermal averaging}
\label{thermal}

We define the thermal averaging by denoting that a function $A(\epsilon)$
should take its value onto the Fermi level for $T=0$
\be
A(\epsilon_f)&=&\int d\epsilon \,\delta(\epsilon-\epsilon_f)A(\epsilon)
\nonumber\\
&\approx&\frac{1}{T}\int d\epsilon \,f^0(1-f^0)A(\epsilon)\nonumber\\
&=&\frac{1}{n_D}\int\frac{d^3p}{(2\pi)^3}f^0(1-f^0)A(\epsilon)
\ee
with the density of state $n_D=mp_fT/(2\pi^2)$ for
low temperatures.

%---------------------------------------------------------------

\section{Calculation of Collision integral}
\label{integral}

Performing the collision integral about Fermi functions 
in (\ref{fermiB}) with the dimensionless variables $x=(\epsilon-\mu)/T$ and
$\lambda=\mu/T$ we get 
\be
I^f_B=&T^3&\int\limits^{\infty}_{-\lambda} dx_1\,dx_2\,dx_3\,dx_4\,
\delta(\bigtriangleup x)f^0_{x_1}f^0_{x_2}\bar{f}^0_{x_3}\bar{f}^0_{x_4}
                                                             \label{fermiBx}
\ee
where $f^0_x=(e^{x}+1)^{-1}$, $\bar{f_x^0}=1-f_x^0$ 
and $\bigtriangleup x= x_1+x_2-x_3-x_4$.
The integral (\ref{fermiBx}) can be done exactly using 
the standard identity over 
Fermi functions\footnote{here: $\lambda=\frac{\mu}{T} \gg 0$} \cite{SMJ89}
and we find 
\be
I_B^f&=&\frac{T^3}{2}\int\limits^{\infty}_{-\lambda} dx_1\,
                           f^0_{x_1}\bar{f}^0_{x_1}(\pi^2+x_1^2)\nonumber\\
&=&\frac{2\pi^2}{3}T^3\,.\label{intB1}
\ee
We perform the Fermi integral in (\ref{fermiM}) as above
\be
I_m^f(\Omega)=\frac{I_m(\Omega)+I_m(-\Omega)}{2} \label{doppelI}
\ee
and have for $I_m$
\be
I_m(\pm\Omega)&=&T^3\int\limits^{\infty}_{-\lambda} dx_1\,dx_2\,dx_3\,dx_4\,
\delta(\bigtriangleup x\pm\Omega)\nonumber\\
&&\times\left[f^0_{x_1}\bar{f}^0_{x_1}\left(\bar{f}^0_{x_2}f^0_{x_3}f^0_{x_4}
+f^0_{x_2}\bar{f}^0_{x_3}\bar{f}^0_{x_4}\right)\right],\label{memI1}
\ee
where $\Omega=\omega/T$. 
Writing (\ref{memI1}) in the form
\be
I_m(\pm\Omega)&=&T^3\int\limits^{\infty}_{-\lambda} dx_1\,dx_2\,dx_3\,dx_4\,
\delta(\bigtriangleup x\pm\Omega)\nonumber\\
&&\times \left[f^0_{x_1}f^0_{x_2}\bar{f}^0_{x_3}\bar{f}^0_{x_4}+
               {f^0_{x_1}}^2f^0_{x_2}\bar{f}^0_{x_3}\bar{f}^0_{x_4}
         \left({\rm e}^{\mp\Omega}-1\right)\right],        \nonumber\\ \label{memI2}
\ee
one gets for $\Omega\rightarrow 0$ the result (\ref{intB1}). 
Applying the standard identity\cite{SMJ89} in (\ref{memI2}) we have
\be
I_m(\pm\Omega)&=&\frac{T^3}{2}\int\limits_{-\lambda}^{\infty}dx_1\,
\left[f_{x_1}^0+{f_{x_1}^0}^2\left({\rm e}^{\mp\Omega}-1\right)\right]\nonumber\\
&&\times \bar{f}^0_{x_1\pm\Omega}\left[\pi^2+(x_1\pm\Omega)^2\right]\nonumber\\
&=&T^3\left(\frac{2}{3}\pi^2+\frac{\Omega}{2}^2\right) 
\ee
so that the final result for $I_m^f$ in (\ref{doppelI} ) reads 
\be
I_m^f(\omega)&=&T^3\left[
\frac{2}{3}\pi^2+\frac{1}{2}\left(\frac{\omega}{T}\right)^2\right]\nonumber\\
&=&I_B^f\left[1+\frac{3}{4}\left(\frac{\omega}{\pi T}\right)^2\right]
\ee

\subsection{Without thermal averaging}
\label{nointegral}

Calculating the Fermi integral in (\ref{fermiM}) without the thermal 
averaging (app. \ref{thermal}) we start from (\ref{memI1}) and get 
\be
{\tilde I}_m(\pm\Omega)&=&T^2\int\limits^{\infty}_{-\lambda} dx_2\,dx_3\,dx_4\,
\delta(\bigtriangleup x\pm\Omega)\nonumber\\
&&\times\left(\bar{f}^0_{x_2}f^0_{x_3}f^0_{x_4}
+f^0_{x_2}\bar{f}^0_{x_3}\bar{f}^0_{x_4}\right).\label{nomemI1}
\ee
Rewriting (\ref{nomemI1}) in the form
\be
{\tilde I}_m(\pm\Omega)&=&T^2\int\limits^{\infty}_{-\lambda} dx_2\,dx_3\,dx_4\,
\delta(\bigtriangleup x\pm\Omega)\nonumber\\
&&\times f^0_{x_2}\bar{f}^0_{x_3}\bar{f}^0_{x_4}
           \left({\rm e}^{-x_1\mp\Omega}+1\right),\label{memI3}
\ee
we apply the standard identity\cite{SMJ89} and find
for ${\tilde I}_m$ 
\be
{\tilde I}_m(x_1,\pm\Omega)&=&
\frac{T^2}{2\,\,}\bar{f}^0_{x_1\pm\Omega}\left[\pi^2+(x_1\pm\Omega)^2\right]
                                      \left({\rm e}^{-x_1\mp\Omega}+1\right).\nonumber\\
&&
\ee
Since $\epsilon_f\approx\mu$ we have $x_1\rightarrow 0$ and get for ${\tilde I}^f_M$
(\ref{doppelI})  
\be
{\tilde I}_m^f(\omega)=
        \frac{T^2\pi^2}{2}\left[1+\left(\frac{\omega}{\pi T}\right)^2\right].
\ee
The resulting relaxation time ${\tilde \tau}_m$ reads finally
\be
 \frac{1}{{\tilde\tau}_{m}(\omega)}=\frac{3}{4}\frac{1}{\tau_B}\left[1+
  \left(\frac{\omega}{\pi T} \right)^2 \right].
\ee
\bibliography{refer}

\begin{thebibliography}{10}

\bibitem{SBT91}
A. Smerzi, A. Bonasera, and M. Di~Toro, Phys. Rev. C {\bf 44},  1713  (1991).

\bibitem{ABA95}
S. Ayik, M. Belkacem, and A. Bonasera, Phys. Rev. C {\bf 51},  611  (1995).

\bibitem{KPS95}
V. Kolomietz, V. Pluiko, and S. Shlomo, Phys. Rev. C {\bf 52},  2480  (1995).

\bibitem{BAB95}
M. Belkacem, S. Ayik, and A. Bonasera, Phys. Rev. C {\bf 52},  2499  (1995).

\bibitem{SUO96}
S. Suomij{\"a}rvi {\it et~al.}, Phys. Rev. C {\bf 53},  2258  (1996).

\bibitem{MOT96}
K. Morawetz and M. Di~Toro, Phys. Rev. C {\bf 54},  833  (1996).

\bibitem{KPS96}
V. Kolomietz, V. Plujko, and S. Shlomo, Phys. Rev. C {\bf 54},  3014  (1996).

\bibitem{TKL97}
M. Di~Toro, V. Kolomietz, and A. Larionov,  in {\em proceedings of the Dubna
  conference on heavy ions} (unpublished, Dubna, 1997).

\bibitem{KTO96}
V. Kondratyev and M. Di~Toro, Phys. Rev. C {\bf 53},  2176  (1996).

\bibitem{HNP96}
E. Hern\'andez, J. Navarro, A.Polls, and J. Ventura, Nucl. Phys. A {\bf 597},
  1  (1996).

\bibitem{RAP96}
E. Ramkrishnan {\it et~al.}, Phys. Rev. Lett. {\bf 76},  2025  (1996).

\bibitem{RAM96}
E. Ramkrishnan {\it et~al.}, Nucl. Phys. A {\bf 549},  49  (1996).

\bibitem{SWA91}
J. Speth and J. Wambach,  in {\em International Review of Nuclear Physics}
  (World Scientific, Singapore, 1991), Chap.~1.

\bibitem{SRS97}
V. Sokolov, I. Rotter, D. Savin, and M. M{\"u}ller, Phys. Rev. C {\bf 56},
  1031  (1997).

\bibitem{ATS94}
V. Abrosimov, M. Di~Toro, and A. Smerzi, Z. Phys. A {\bf 347},  161  (1994).

\bibitem{KAM69}
S. Kamerdzhiev, Yad. Fiz. {\bf 9},  324  (1969).

\bibitem{KLT97}
V. Kolomietz, A. Larionov, and M. Di~Toro, Nucl. Phys. A {\bf 613},  1  (1997).

\bibitem{BRI88}
D. Brink, Nucl. Phys. A {\bf 482},  205  (1986).

\bibitem{BDT86}
D. Brink, A. Dellafiore, and M. Di~Toro, Nucl. Phys. A {\bf 456},  205  (1986).

\bibitem{DMA86}
A. Dellafiore and F. Matera, Nucl. Phys. A {\bf 460},  265  (1986).

\bibitem{BAR96}
V. Baran {\it et~al.}, Nucl. Phys. A {\bf 599},  29  (1996).

\bibitem{CTG93}
M. Colonna {\it et~al.}, Phys. Rev. B {\bf 307},  293  (1993).

\bibitem{RSC82}
P. Ring and P. Schuck, {\em The Nuclear Many Body Problem} (Springer, Berlin,
  1982).

\bibitem{BBB83}
G. Bertsch, P. Bortignon, and R. Broglia, Rev. of Mod. Phys. {\bf 55},  287
  (1983).

\bibitem{PNO68}
D. Pines and P. Nozi{\`e}res, {\em The Theory of Quantum Liquids}
  (Addison-Wesley, New York, 1968), Vol.~1.

\bibitem{KBA62}
L. Kadanoff and G. Baym, {\em Quantum Statistical Mechanics} (Addison-Wesley,
  New York, 1962).

\bibitem{LSV86}
P. Lipavsk\'y, V. {\v S}pi{\v c}ka, and B. Velick\'y, Phys. Rev. B {\bf 6},
  3189  (1986).

\bibitem{M94}
K. Morawetz, Phys. Lett. A {\bf 199},  241  (1995).

\bibitem{MSL97}
K. Morawetz, V. \v{S}pi\v{c}ka, and P. Lipavsk{\'y}, Phys. Rev. Lett.  (1997),
  submitted.

\bibitem{LIN54}
J. Lindhard, Kgl. Danske Videnskab. Selskab. Mat. Fys. Medd. {\bf 28},  8
  (1954).

\bibitem{BRV94}
F. Braghin and D. Vautherin, Phys. Lett. B {\bf 333},  289  (1994).

\bibitem{VBR72}
D. Vautherin and D. Brink, Phys. Rev. C {\bf 5},  626  (1972).

\bibitem{SJE50}
H. Steinwedel and J. Jensen, Z. f. Naturforschung {\bf 5},  413  (1950).

\bibitem{KKE86}
W.-D. Kraeft, D. Kremp, W. Ebeling, and G. R{\"o}pke, {\em Quantum Statistics
  of Charged Particle Systems} (Akademie-Verlag, Berlin, 1986).

\bibitem{LPI79}
E. Lifschitz and L. Pitajewski, {\em Lehrbuch der Theoretischen Physik} (Nauka,
  Moskau, 1978), Vol.~10.

\bibitem{SMJ89}
H. Smith and H. Jensen, {\em Transport Phenomena} (Clarendon Press, Oxford,
  1989).

\bibitem{BPE91}
G. Baym and C. Pethick, {\em Landau Fermi-Liquid Theory} (Wiley, New York,
  1991).

\bibitem{ABO92}
S. Ayik and D. Boilley, Phys. Rev. B {\bf 276},  263  (1992), errata Phys.
  Lett. B {\bf 284}, 482 (1992).

\bibitem{MER70}
N. Mermin, Phys. Rev. B {\bf 1},  2362  (1970).

\bibitem{HPR93}
H. Heiselberg, C.~J. Pethick, and D.~G. Ravenhall, Ann. Phys. {\bf 223},  37
  (1993).

\bibitem{GRY94}
I. Gradshteyn and I. Ryzhik, {\em Table of Integrals, Series, and Products}
  (Academic Press, San Diego, 1994).

\bibitem{MFW97}
K. Morawetz, R. Walke, U. Fuhrmann, and M. Di~Toro, Phys. Rev. C  (1997),
  submitted.

\bibitem{BER88}
S. Dietrich and B. Berman, Nucl. Data Tabl. {\bf 38},  199  (1988).

\bibitem{DGB96}
P. Donati, N. Giovanardi, P. Bortignon, and R. Brogila, Phys. Lett. B {\bf
  383},  15  (1996).

\bibitem{DBB96}
P. Donati, P. Bortignon, and R. Brogila, Z. Phys. A {\bf 354},  249  (1996).

\bibitem{LWI72}
D. Lengreth and G. Wilkins, Phys. Rev. B {\bf 34},  6933  (1972).

\bibitem{SLI95}
V. {\v S}pi{\v c}ka and P. Lipavsk\'y, Phys. Rev. B {\bf 52},  14615  (1995).

\bibitem{BBS94}
M. Bonitz {\it et~al.}, Phys. Rev. E {\bf 49},  5535  (1994).

\end{thebibliography}
\bibliographystyle{prsty}

\end{document}